\documentclass[prb,twocolumn,showpacs,preprintnumbers,amsmath,amssymb]{revtex4}

\usepackage{graphicx}
\usepackage{dcolumn}
\usepackage{bm}
\usepackage{color}

\def\be{\begin{equation}}       \def\ee{\end{equation}}
\def\bea{\begin{eqnarray}}      \def\eea{\end{eqnarray}}
\def\bes{\begin{subequations}}  \def\ees{\end{subequations}}
\def\half{\frac{1}{2}}
\def\dag{\dagger}
\def\non{\nonumber}
\def\mbz{\text{MBZ}}
\def\k{_{\bf k}}
\def\hc{\text{h.c.}}

\begin{document}

\title{Varied Perturbation Theory for the Dispersion Dip in the Two-Dimensional
Heisenberg Quantum Antiferromagnet}

\author{G\"{o}tz S. Uhrig}
\email{goetz.uhrig@tu-dortmund.de}
\affiliation{Lehrstuhl f\"{u}r Theoretische Physik I, Technische Universit\"{a}t Dortmund,
Otto-Hahn Stra{\ss}e 4, 44221 Dortmund, Germany}

\author{Kingshuk Majumdar}
\affiliation{Department of Physics, Grand Valley State University, Allendale, 
Michigan 49401, USA}
\email{majumdak@gvsu.edu}
\date{\today}

\begin{abstract}
\label{abstract}
We study the roton-like dip in the magnon dispersion at the boundary of the Brillouin zone 
in the isotropic $S=1/2$ Heisenberg quantum antiferromagnet. This high-energy feature is 
sometimes seen as indication of a fractionalization of the magnons to spinons. In this article,
 we provide evidence that the description of the dip in terms of magnons can be improved significantly
by applying more advanced evaluation schemes. In particular, we illustrate the
usefulness of the application of the principle of minimal sensitivity in varied perturbation
theory. Thereby, we provide an example for the application of this approach 
to an extended condensed matter problem governed by correlations which can trigger
analogous investigations for many other systems.
\end{abstract}



\maketitle

\section{\label{sec:Intro}Introduction}

Quantum antiferromagnetism is a long-standing issue which cannot
be reviewed briefly, but see for instance Ref.\ \onlinecite{manou91}.
Yet there are still important aspects which are not clear. In particular,
the dynamics at high energies is not yet fully understood. 
But the quantitative understanding of the high-energy dynamics is
of increasing experimental relevance. For instance, the dispersion of
spin waves in the two-dimensional parent compounds of the high-temperature
superconductors displays dips which can only be accounted for by 
considering subdominant exchange couplings as well, for references
and further discussion see  Ref.\ \onlinecite{majum12a}. 
The relevance of the magnetic dynamics in the total Brillouin zone 
including the high-energy behavior close to the Brillouin zone 
boundary has recently been
emphasized by inelastic X-ray scattering for doped and undoped
cuprates \cite{letac11}.

The purpose of the present article is partly methodological.
Thus we discuss the fundamental system, namely the isotropic Heisenberg
quantum antiferromagnet with spin $S=1/2$ and nearest neighbor exchange 
$J>0$
\be
H =\half J \sum_{i,\delta}{\bf S}_{i} \cdot {\bf S}_{i+\delta},
\label{eq:hamilton1}
\ee
where the subscript $i$ runs over all sites and $\delta$ runs
over the vectors to the adjacent sites. This model is very well
studied. But we focus on the roton-like dip at the wave vector
${\bf k}=(\pi,0)$ and its equivalent values (the lattice spacing is set to unity).
This feature represents an open issue because it eludes
precise calculation within spin wave theory.

On the one hand,
the dispersion $\omega({\bf k})/S$ at ${\bf k}=(\pi,0)$ and at
${\bf k}=(\pi/2,\pi/2)$ takes precisely the same value in linear 
spin wave theory which represents the leading order in an expansion in $1/S$,
where $S$ is the spin value. In the next-leading,
first order this remains true as well. On the other hand,
series expansion for $S=1/2$ around the Ising limit predicts a dip of 
8.6\% \cite{singh95,zheng05} which is confirmed by quantum Monte Carlo 
with a dip of 9.6\% \cite{sandv01}. 

The idea suggests itself that further corrrections of spin wave theory
in $1/S$ cure the above discrepancy \cite{igara92,hamer92,igara05}. 
But this seems not to be the case.
Although in second order $1/S^2$, a small dip appears, it takes only 1.4\%
which is far from what one would like to have \cite{syrom10}.
This number improves by about a factor of 2 upon passing to the third order $1/S^3$ to 3.2\%.
But this is still far from the series and quantum Monte Carlo result \cite{syrom10}.
The convergence turns out to be particularly slow.

One may view this observation as a mere mathematical problem. But one may
also wonder why the convergence is so slow and come to the conclusion that
the underlying physical description in terms of spin waves, also called magnons, 
is not appropriate and that the true nature of the elementary excitations is a 
different one, for instance that the magnons disintegrate to spinons.
Such a view is indeed discussed in the interpretation of the experimental findings
\cite{ronno01,chris04b,chris07,headi10}. In Ref.\ \onlinecite{chris07} the dip is given to be 
7(1)\% analysing the experimental data. The naive analysis of the peak
positions in Figs.\ 3a and 3b in Ref.\ \onlinecite{chris07} suggests even 
about 10\% for the dip. In any case, a sizable dip is an experimentally
well-supported fact.

Concerning the issue of the elementary excitations, it is useful to recall 
a well-studied system where the same question was discussed. In Heisenberg $S=1/2$ ladders
with two legs the elementary excitations are $S=1$ triplons because no long-range
order occurs. But multi-particle continua are sizeable as well \cite{knett01b,schmi05b}
and they may be taken as precursors of a fractionalization towards spinons \cite{knett01b}.
Interestingly, the important multi-particle continua and their energetic vicinity to 
the dispersion of the elementary triplons \cite{schmi05b} induces a dip in their dispersion
at $k=0$ compared to $k=\pi/2$. But high order perturbative results are necessary to capture
this effect \cite{knett00a,trebs00,knett01b}.

These observations led us to look for a quantitative description of the dip
in the dispersion on the square lattice in terms of magnons. Since standard
perturbation theory seems to be not particularly efficient, 
see above, we follow a modified approach. The basic idea is to stick essentially to  
a second order perturbative approach, but to vary the starting point of the
perturbation. This means that we vary the unperturbed Hamiltonian $H_0$ in order
to improve the results. Of course, the total Hamiltonian $H=H_0+H_P$ may not 
be changed so that a variation of  $H_0$ will automatically imply a variation of
the perturbing part $H_P$.

Let us assume that $H_0$ depends on some parameter $u$ or on a set of parameters
$\vec{u}$ which we may vary. How does one determine the appropriate starting point
$H_0(u_0)$ for the perturbation? In this issue we follow the principle of minimal
sensitivity \cite{steve81}. The underlying idea is that the exact diagonalization of $H$
does not depend on the starting point $u_0$. This, of course, will not be true
for a generic approximative scheme. Thus a quantity such as the ground state energy $E_0$
will depend on $u$ if computed approximately: $E_0^\text{appr}(u)$.
Then, the principle of minimal sensitivity suggests to choose $u_0$ such that
$E_0^\text{appr}(u)$ depends minimally on $u$ in the vicinity of $u_0$.
Thus, one should choose local extrema or saddle points to determine the starting value
$u_0$. Given that $E_0^\text{appr}(u)$ is differentiable one obtains as
defining equation 
\be
\label{eq:variation}
\partial_u E_0^\text{appr}(u)\Big|_{u=u_0}=0 .
\ee
Note that for $n$ parameters $u_i$ the above prescription
implies $n$ equations $
\partial_{u_i} E_0^\text{appr}(\vec{u})|_{\vec{u}=\vec{u}_0}=0$
to determine $\vec{u}_0$. In case that $E_0^\text{appr}(\vec{u})$ is
not differentiable at the points of interest we look for local extrema or
saddle points. This is analogous to standard thermodynamics where the physical phase
is represented by a local extrema or saddle points of a thermodynamic potentials. Cusps may occur as well
and they generically indicate first order transitions. This exemplifies that 
the non-differentiability does not invalidate the prescription to look for 
local extrema or saddle points.

If the perturbation is performed around a bilinear bosonic Hamiltonian
the dependence $H_0(u)$ may equivalently be replaced by the
dependence of the set $\{a_i(u)\}$ of annihilation operators and the
corresponding creation operators which diagonalize $H_0(u)$.
This approach has been used to illustrate the usefulness of
the principle of minimal sensitivity in perturbative calculations
and for continuous unitary transformations \cite{steve81,dusue04a}.
Below, we will use it to improve perturbative spin wave calculations 
for the Heisenberg model on a square lattice.
The spin operators will be represented as introduced by Dyson and Maleev
\cite{dyson56a,dyson56b,malee57,malee58}
so that the  bosonic approach can be directly put to use.

The article is set up as follows. In the following section II, we introduce the model and its
bosonic representation. In particular, the variation of the bosonic
description will be explained. In Sect.\  III we present results for the variation
in two parameters. Results for the ground state energy and for the dispersion
are shown. Finally, the article is concluded in Sect.\ IV.

\section{\label{sec:modelmethod}Model and Method}

\subsection{Model}

Expressed in the usual spin operators the Hamiltonian is given 
in Eq.\ \eqref{eq:hamilton1}. Exploiting that the square lattice is
bipartite we may write
\be
H =J \sum_{i\in\Gamma_A,\delta}{\bf S}^A_{i} \cdot {\bf S}^B_{i+\delta},
\label{eq:hamilton2}
\ee
where $\Gamma_A$ is the lattice made only from all $A$ sites.
Finally, we will focus on $S=1/2$. But for introducing
the bosonic representation it is convenient to treat general spin.
We use the Dyson-Maleev representation \cite{dyson56a,dyson56b,malee57,malee58}
\begin{subequations}
 \label{eq:dyson} 
\begin{eqnarray}
S_{Ai}^+ &=& \sqrt{2S}\Big[a_i-  \frac {a_i^\dag a_i a_i}{(2S)}\Big],\;
S_{Ai}^-= \sqrt{2S}a_i^\dag,\nonumber\\
S_{Ai}^z &=& S-a^\dag_ia_i,  \\ 
S_{Bj}^+ &=& \sqrt{2S} \Big[b_j^\dag- \frac {b_j^\dag b_j^\dag b_j}{(2S)} \Big],\;
S_{Bj}^- = \sqrt{2S} b_j,\nonumber\\
S_{Bj}^z &=& -S+b^\dag_j b_j,
\end{eqnarray}
\end{subequations}
where $a_i^{(\dag)}$ are bosonic creation/annihilation operators on the $A$-sites
and $b_i^{(\dag)}$ on the $B$-sites. Next, we transform these bosonic operators in
momentum space. We stress that the momenta ${\bf k}$ are taken from the magnetic 
Brillouin zone (MBZ), which is a tilted square in $k$-space with the corners
$(\pm\pi,0)$ and $(0,\pm\pi)$, because the real space coordinate $i$ runs 
over $\Gamma_A$.

\subsection{Basic Steps}

Next, we perform a conventional Bogoliubov transformation 
respecting translational invariance
\be
a^\dag_{\bf k} =l_{\bf k} \alpha_{\bf k}^\dag + m_{\bf k}\beta_{-{\bf k}},\;\;\;
b_{-\bf k} =m_{\bf k} \alpha_{\bf k}^\dag + l_{\bf k}\beta_{-{\bf k}},
\ee
where $\alpha_{\bf k}^{(\dag)}$ and $\beta_{\bf k}^{(\dag)}$ are the new operators
in which we express the Hamiltonian. The prefactors $l_{\bf k}$ and
$m_{\bf k}$ can be chosen at will as long as they fulfil $l_{\bf k}^2+ m_{\bf k}^2=1$
where we assume them to be real. The freedom of choice for these prefactors
provides us with the possibility to choose  the starting point of the perturbation
theory as described in the Introduction.
Below, in Sect.\ \ref{ssc:variation},
we will specify how $l_{\bf k}$ and $m_{\bf k}$ depend on the variational
parameters.

We find it convenient to parametrize the prefactors by
\bea
l_{\bf k} &=& \Big[\frac {1+\mu_{\bf k}}{2\mu_{\bf k}} \Big]^{1/2},\;\;
m_{\bf k} = -\Big[\frac {1-\mu_{\bf k}}
{2\mu_{\bf k}} \Big]^{1/2}=: -x_{\bf k}l_{\bf k},\nonumber\\
x_{\bf k} &=& \Big[\frac {1-\mu_{\bf k}}
{1+\mu_{\bf k}} \Big]^{1/2},
\eea
where $\mu_{\bf k}$ can still be chosen freely as long as $|\mu_{\bf k}|\le 1$ holds.
To elucidate the above parametrization we recall that the choice
\be
\mu_{\bf k}=\sqrt{1-\gamma_{\bf k}^2} \;\;\; \gamma_{\bf k}:=\frac{1}{2}(\cos(k_x)+\cos(k_y))
\ee
leads to the standard linear spin wave description. We will come back to this point
in Sect.\ \ref{ssc:variation} where we will specify how $\mu_{\bf k}$ is modified
as funciton of variational parameters.

First, however, we express the Hamiltonian in the fields 
$\alpha_{\bf k}^{(\dag)}$ and $\beta_{\bf k}^{(\dag)}$.
We split it according to
\be
\label{eq:hamilton3}
H=H_\text{cl}+H_\text{bl}+H_\text{ql} , 
\ee
where $H_\text{cl}=-4JS^2N$ simply stands for the classical ground state energy; note
that here $N$ is the number of $A$-sites.
The second term $H_\text{bl}$ stands for the part which stems from the bilinear
terms if $H$ is expressed in the original bosonic fields in \eqref{eq:dyson}.
It reads
\be
H_\text{bl} = E_{01} + H_\text{D1}+ H_\text{B1},
\label{eq:ham_bl}
\ee
where
\begin{subequations}
\bea
E_{01} &:=& 8JS \sum_{{\bf k}\in\mbz} l^2\k x\k(x\k-\gamma\k)\\
H_\text{D1} &:=& 4JS \sum_{{\bf k}\in\mbz} 
A_{1\bf k}(\alpha\k^\dag\alpha\k + \beta\k^\dag\beta\k) \\
H_\text{B1}  &:=& 4JS \sum_{{\bf k}\in\mbz}
B_{1\bf k}(\alpha\k^\dag\beta_{-\bf k}^\dag+\hc).
\eea
\end{subequations}
The momentum dependent prefactors are given by
\begin{subequations}
\bea
A_{1\bf k} &:=& l^2\k(1-2x\k\gamma\k+x^2\k)
\\
B_{1\bf k} &:=& l^2\k(\gamma\k-2x\k+\gamma\k x^2\k).
\eea
\end{subequations}
Doing the same for the quartic part $H_\text{ql}$ yields
\be
H_\text{ql} = E_{02} + H_\text{D2}+ H_\text{B2}+ H_\text{V},
\label{eq:ham_ql}
\ee
with
\begin{subequations}
\bea
E_{02} &:=& -JN A^2_2
\\
\label{eq:A2}
A_2 &:=& \frac{2}{N}\sum_{{\bf k}\in\mbz} l^2\k(x\k\gamma\k-x^2\k)
\\
H_\text{D2} &:=& 2J \sum_{{\bf k}\in\mbz} 
A_{2\bf k}(\alpha\k^\dag\alpha\k + \beta\k^\dag\beta\k) 
\\
H_\text{B2}  &:=& 2J \sum_{{\bf k}\in\mbz}
B_{2\bf k}(\alpha\k^\dag\beta_{-\bf k}^\dag+\hc),
\eea
\end{subequations}
where we find
\begin{subequations}
\bea
A_{2\bf k} &=& A_2\cdot A_{1\bf k}
\\
B_{2\bf k} &=& A_2\cdot B_{1\bf k}.
\eea
\end{subequations}
Of course, the simplicity of the last relation results from the
simplicity of the original model which is characterized only by
nearest neighbor couplings which are all renormalized by the
mean-field effects in the same way.

The quadrilinear interaction part is given by
the normal-ordered expression
\bea
H_\text{V} &=&  -  \frac {J}{N}\sum_{1234}
\delta_{12}^{34}\; l_1l_2l_3l_4
\Big[V_{1234}^{(1)} \alpha_1^\dag \alpha_2^\dag \alpha_3 \alpha_4 
\non\\
&+&2V_{1234}^{(2)}\alpha_1^\dag \beta_{-2}\alpha_3 \alpha_4 
+ 2V_{1234}^{(3)}\alpha_1^\dag \alpha_2^\dag \beta_{-3}^\dag \alpha_{4}
\non \\
&+& 4V_{1234}^{(4)}\alpha_1^\dag \alpha_3 \beta_{-4}^\dag \beta_{-2}
+ 2V_{1234}^{(5)}\beta_{-4}^\dag \alpha_3 \beta_{-2} \beta_{-1}
\non\\
&+& 2V_{1234}^{(6)}\beta_{-4}^\dag \beta_{-3}^\dag \alpha_{2}^\dag \beta_{-1} 
+ V_{1234}^{(7)}\alpha_1^\dag \alpha_2^\dag \beta_{-3}^\dag \beta_{-4}^\dag
\non \\
&+& V_{1234}^{(8)}\beta_{-1} \beta_{-2}\alpha_3 \alpha_4
+V_{1234}^{(9)}\beta_{-4}^\dag \beta_{-3}^\dag \beta_{-2} \beta_{-1}\Big],\quad
\eea
where the subscripts $i=1,2,3,4$ stand for the momenta 
${\bf k}_i$ and  $-i$ stands for $-{\bf k}_i$.
The conservation of momentum in the lattice is ensured by the
Kronecker symbol $\delta_{12}^{34}$ which
implies ${\bf k}_1+{\bf k}_2={\bf k}_3+{\bf k}_4$ modulo reciprocal
lattice vectors from the reciprocal lattice $\Gamma_A^*$
of the $A$-sites, i.e., ${\bf g}\in\Gamma_A^*$ means
${\bf g}=(n\pi,m\pi)$ with the integers $n, m$ if the lattice constant
of the original square lattice is set to unity.
The vertex functions  $V_{1234}^{(i)}$ are given explicitly in App.\ \ref{sec:vertex}.

Now we can combine the diagonal parts in
\bes
\bea
\label{eq:diagonal}
H_D &:=& E_{00} + H_\text{D1}+ H_\text{D2}
\\
 &=& E_{00} + 4J\sum_{{\bf k}\in\mbz} \omega_{\bf k} (\alpha\k^\dag\alpha\k + \beta\k^\dag\beta\k) 
 \\
  E_{00} &:=&H_\text{cl}+E_{01}+E_{02}
 \\
  &=& -4JS^2N-J(4SA_2+A_2^2)N
 \\
 \omega_{\bf k}  &=& (S+\frac{1}{2}A_2)l^2\k(1-2x\k\gamma\k+x^2\k)
\eea
\ees
and the perturbing part $H_P$ in
\bes
\bea
\label{eq:perturb}
H_P &:=& H_B +H_V
\\
H_B &:=& 4J\sum_{{\bf k}\in\mbz} B\k(\alpha\k^\dag\beta_{-\bf k}^\dag+\hc)
\\
B\k &:=& (S+\frac{1}{2}A_2) l^2\k(\gamma\k-2x\k+\gamma\k x^2\k)
\eea
\ees
where $H_B$ in \eqref{eq:perturb} stems from the sum $H_\text{B1}+ H_\text{B2}$.

\subsection{Approximate Evaluation}

A straightforward procedure is to use standard perturbation theory in $H_V$, for instance in 
second order, to compute the ground state energy $E_0$ and the dispersion $\omega({\bf k})$
in an approximate way. (Note the difference between $\omega\k$, the dispersion in the 
unperturbed Hamiltonian $H_D$ and the dispersion $\omega({\bf k})$ of the full Hamiltonian.)
First, we focus on the ground state energy because its local saddle point
\eqref{eq:variation} will determine $\{\mu\k\}$. The correction $\Delta E_B$ due to $H_B$ can 
be easily computed to infinite order in $B\k$ analytically by Bogoliubov transformation
\be
\Delta E_B = -2JN(2S+A_2)({\cal A}_2-A_2),
\ee
where ${\cal A}_2$ is the value for $A_2$ if we diagonalize the bilinear Hamiltonian
from the very beginning, i.e., ${\cal A}_2=A_2$ as given by Eq.\ \eqref{eq:A2} for
$\mu\k=\sqrt{1-\gamma\k^2}$.
The correction $\Delta E_V$ involving $H_V$ are much more complicated so that we determine them
only in second order in $H_V$
\be
\Delta E_V = -\frac{J}{N^2} \sum_{1234} 
\frac{\delta_{12}^{34} (l_1l_2l_3l_4)^2 V^{(7)}_{1234} V^{(8)}_{4321}}
{\omega_1+\omega_2+ \omega_3 +\omega_4}
\ee
Formally, there could also be a second order contribution which is linear in $H_B$ 
and in $H_V$, but no such term contributes to the ground state energy. Hence
the approximate ground state energy $E_0$ is given by
\be
\label{eq:E0}
E_0 = E_{00} + \Delta E_B  + \Delta E_V.
\ee

In the same fashion, we compute the dispersion. The influence of $H_B$ is again
taken into account in infinite order yielding
\be
\omega_B({\bf k}) = 2J(2S+A_2)\sqrt{1-\gamma^2\k}.
\ee
The additional second order correction $\Sigma_V({\bf k})$ reads
\bes
\bea
\Sigma_V({\bf k}) &=& \Sigma_{BV}({\bf k}) + \Sigma_{VV}({\bf k})
\\
\Sigma_{BV}({\bf k}) &=& \frac{2J l^2\k}{N}\sum_{\bf p} \frac{l_{\bf p}^2}{\omega_{\bf p}}
B_{\bf p}(V^{(2)}_{{\bf k}{\bf p}{\bf p}{\bf k}} +
V^{(3)}_{{\bf k}{\bf p}{\bf p}{\bf k}})
\\ \non
\Sigma_{VV}({\bf k}) &=& \frac{2J l^2\k}{N^2}\sum_{{\bf p},{\bf q},{\bf s}} (l_{\bf p}l_{\bf q}l_{\bf s})^2
\delta_{{\bf k}{\bf p}}^{{\bf q}{\bf s}}\left[ \frac{V^{(2)}_{{\bf k}{\bf p}{\bf q}{\bf s}} 
V^{(3)}_{{\bf s}{\bf q}{\bf p}{\bf k}}}{\omega_{\bf k}-\omega_{\bf p}-\omega_{\bf q}-\omega_{\bf s}}
\right.
\\
&&\left.
-\frac{V^{(7)}_{{\bf k}{\bf p}{\bf q}{\bf s}} 
V^{(8)}_{{\bf s}{\bf q}{\bf p}{\bf k}}}{\omega_{\bf k}+\omega_{\bf p}+\omega_{\bf q}+\omega_{\bf s}}
\right]
\eea
\ees
so that the total approximate dispersion finally is given by
\be
\omega({\bf k}) = \omega_B({\bf k}) +\Sigma_V({\bf k}).
\ee
Note that in the quadratic correction $\Sigma_{BV}({\bf k})$ both perturbing terms
$H_B$ and $H_V$ enter.

\subsection{Variation of $H_D$}
\label{ssc:variation}

The equations similar to the above can be found in many previous
approaches \cite{igara92,hamer92,igara05,syrom10,majum12a}. 
The main difference is that in the previous equations $\mu\k$
was chosen such that $H_B$ vanished or appeared only in subdominant
orders in $1/S$. The equations above for arbitrary $\mu\k$
are more general. They allow us to vary what we call an
$\alpha\k^\dag$ or $\beta\k^\dag$ excitation. Thereby, the diagonal part of the
perturbation $H_D$ is varied and we can apply the principle of minimal
sensitivity by looking for local saddle points of $E_0$ as it results
from the approximate calculation.

Pursuing this line of argument we should vary $\mu\k$ at each point in
the magnetic Brillouin zone in the range $1\ge |\mu\k|$. This, however,
is far too ambitious because of the macroscopic number of 
parameters to be varied. Thus, to simplify the approach we choose
a particular parametrization of $\mu\k$ which relies only on a small
number of parameters. In the present work, we want to illustrate the
approach in principle and restrict ourselves to two free parameters.
Moreover, it is reasonable to choose $\mu\k$ close to $\sqrt{1-\gamma\k^2}$
which would correspond to the correct solution in the limit $S\to\infty$.
Therefore, our choice is
\be
\label{eq:parametrization}
\mu\k = (1-f\k)\sqrt{1-\gamma\k^2},
\ee
where
\bes
\label{eq:f_def}
\bea
f\k &&:=
\\ \non
&& v\cos(k_x)\cos(k_y)+|v|+u(\cos(k_x)+\cos(k_y)-2)
\eea
for $u\le|v|/4$ and 
\bea
f\k &&:=
\\ \non
&& v\cos(k_x)\cos(k_y)+u(\cos(k_x)+\cos(k_y)+2)
\eea
\ees
otherwise. The above choice is motivated by two arguments.
First, we intend to include the cosine terms which go beyond nearest neighbor
processes. The simplest choice are the two next-nearest neighbor processes 
included above. Second, $f\k$ may not become negative because $|\mu\k|$ can
exceed unity in this case. This is particularly important at
the border of the magnetic Brillouin zone where $\gamma\k=0$.
In addition, $f\k$ may not exceed unity because $\mu\k$ should not change sign.
This implies that $u$ and $v$ may not be chosen too large.

On the boundary, i.e., for $\gamma\k=0$, we choose $k_x=q$ $k_y=\pi-q$ and obtain
for $u\le|v|/4$
\bes
\label{eq:boundary_dispersion}
\be
f_q = (v/2-2u)(1-\cos(2q))+ |v|-v
\ee
and 
$u\ge|v|/4$
\be
f_q = (2u-v/2)(1+\cos(2q)).
\ee
\ees
We see that $f_q\ge 0$ is ensured. Note for future reference that for $u=v/4$ no
dispersion along the boundary of the magnetic Brillouin zone occurs
so that this line is special.

\section{\label{sec:result}Results}

The results presented below are evaluated for $S=1/2$.
First, we analyze the dependence of the ground state energy on the chosen parameters.
It turns out that the most interesting parameter region is $u,v \ge 0$.

\subsection{\label{sec:gse}Ground State Energy}

\begin{figure}[htb]
\centering
\includegraphics[width=\columnwidth,clip]{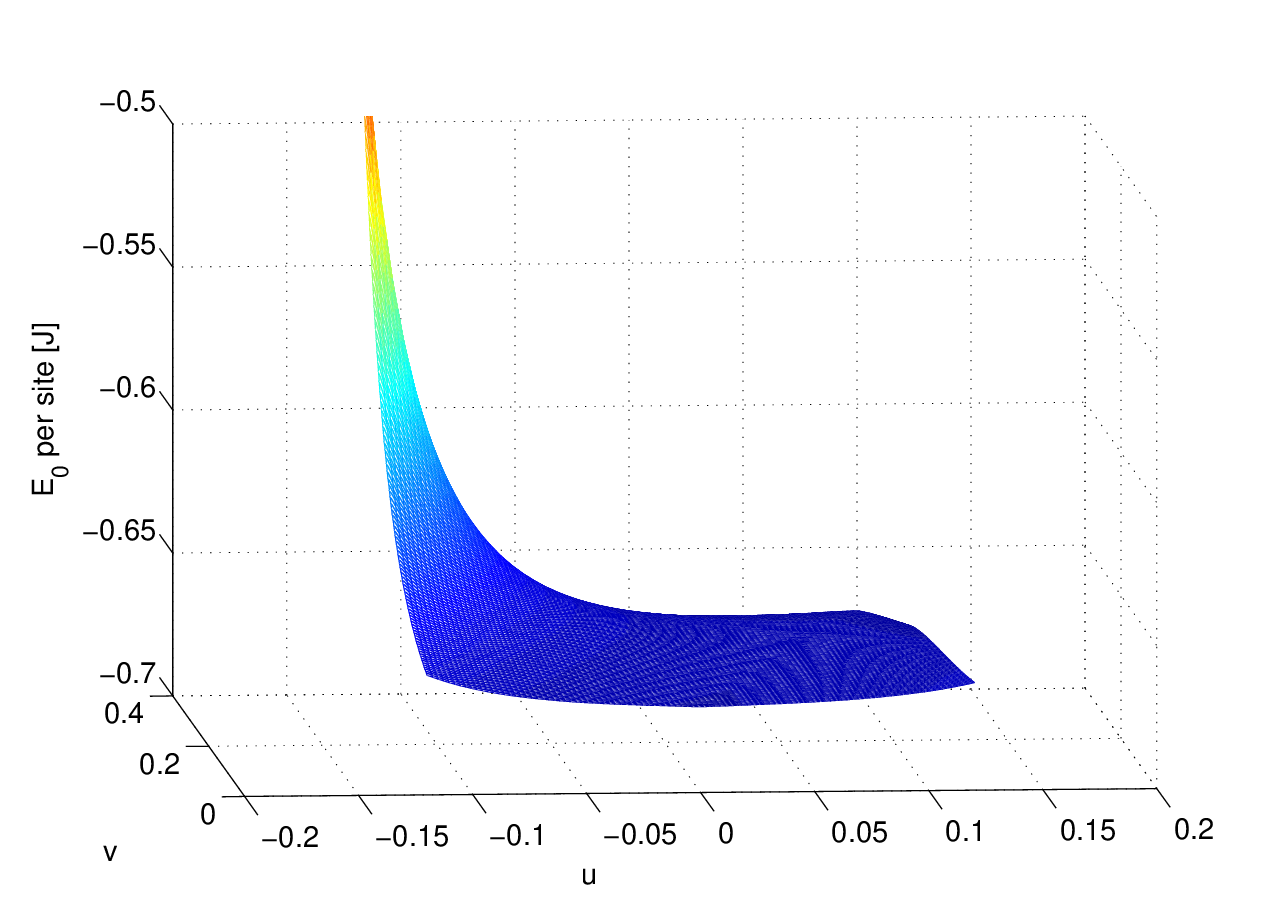}
\includegraphics[width=\columnwidth,clip]{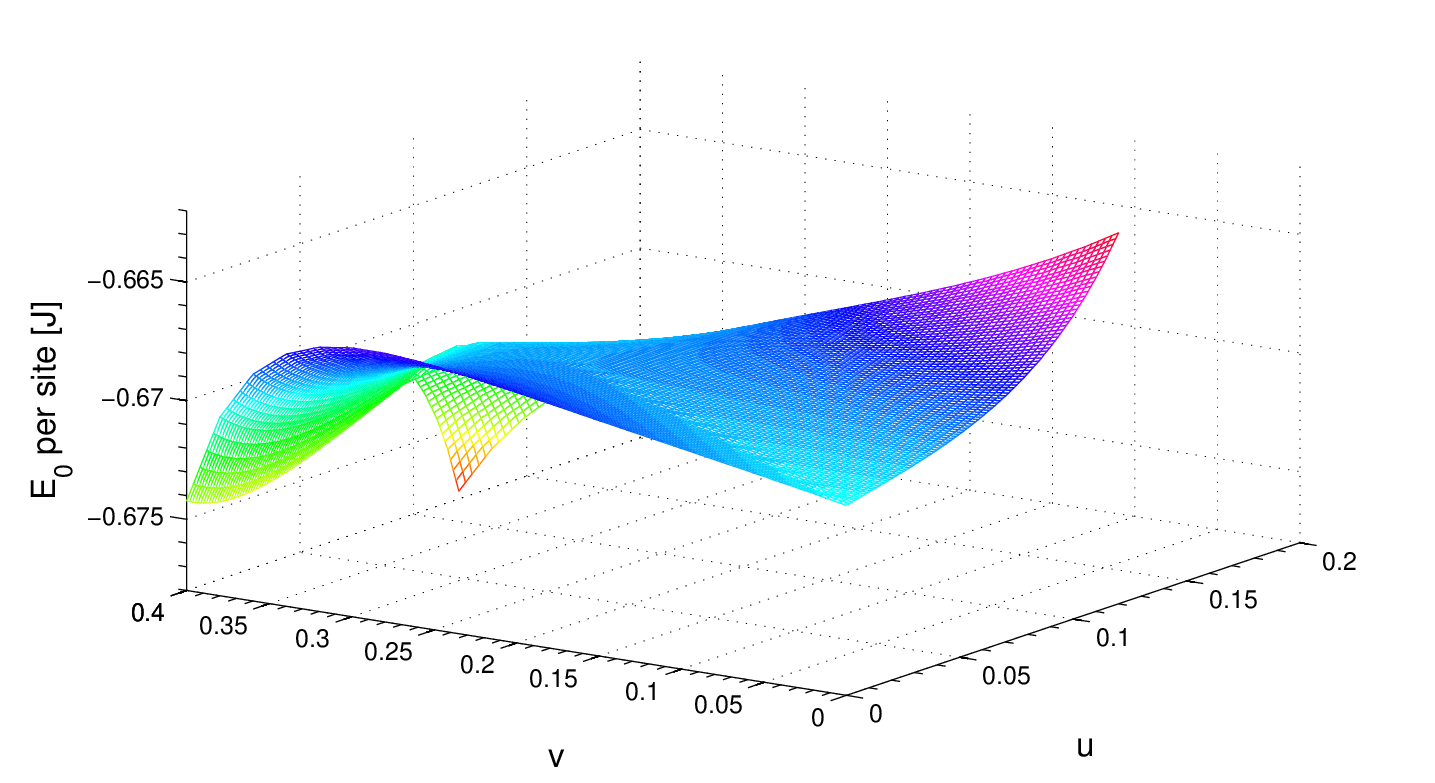}
\caption{\label{fig:energy_landscape} 
(Color online)
Approximate ground state energy $E_0(u,v)$ per site of
the original lattice, i.e., $E_0/(2N)$, as computed by
Eq.\ \eqref{eq:E0}. If the calculation were exact, $E_0$ 
should be constant. Due to the approximations used this is not the
case. The best strategy is to look for local extrema or saddle points
because they represent points where $E_0$ is stationary at least
locally.
Upper panel: Overall view, no extrema or saddle points are discernible.
Lower panel: For $u\ge 0$, the energy landscape displays more structure
and a saddle point can be presumed in the middle of the figure.
Note that for clarity the color coding in the lower panel
is different from the one in the upper panel.
}
\end{figure}

Fig.\ \ref{fig:energy_landscape} shows the energy dependence of $E_0$
as given by \eqref{eq:E0} on the two parameters $u$ and $v$.
Obviously, no dominant local minima or maxima catch our eye
in the upper panel. In the lower panel, one can presume a saddle point
in the center of the figure. Generally, very little dependence
on $u$ and $v$ occurs in the middle region displayed in the lower
panel.

\begin{figure}[htb]
\centering
\includegraphics[width=0.75\columnwidth,clip]{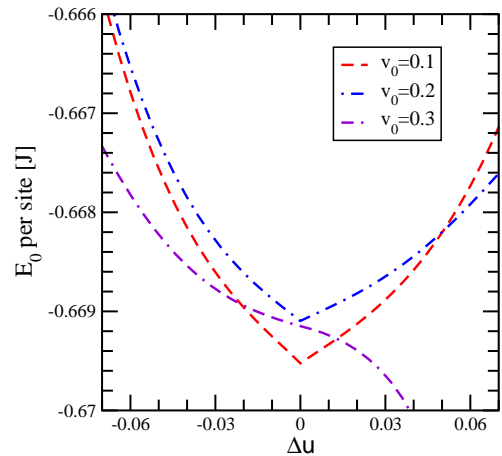}
\caption{\label{fig:cuts_across} 
(Color online)
Cuts of $E_0(u,v)/(2N)$ perpendicular to the line $u=v/4$ along the line
$(u_0+\Delta u,v_0-\Delta u/4)$. The cusps at $\Delta u=0$ for not too large
$v_0$ are obvious.
}
\end{figure}

Closer inspection of this range
of $u$ and $v$, see Fig.\ \ref{fig:cuts_across},
  shows that there is a line of small cusps given by
$u=v/4$ as long as $v$ is not too large,
 see also right panel of Fig.\ \ref{fig:energy_cuts} below.
In view of the definition of $f\k$ in \eqref{eq:f_def} the appearance of such
a cusp may not surprise. In addition, the line $u=v/4$ is special
since it makes any dispersion at the magnetic Brillouin zone boundary vanish.

To elucidate the energy behavior more quantitatively Fig.\ \ref{fig:energy_cuts}
shows two perpendicular cuts through the energy landscape of Fig.\ \ref{fig:energy_landscape}.
The left panel in Fig.\ \ref{fig:energy_cuts} follows the line of cusps along
$u=v/4$. Clearly, a local maximum appears which is located at
$v_0=0.2502(1)$. But the dependence through
this point along a line perpendicular to $u=v/4$ displays the cusp at $v_0$
which is a local minimum. Note, that the definition \eqref{eq:f_def}
is prone to yield cusps as stated above. But it is a priori not clear that
these cusps are extrema in certain directions.
Since the point at $(v_0/4,v_0)$ is a local minimum in one direction, but
a local maximum in the perpendicular direction we are not observing a local extremum, but 
a saddle point though $E_0(u,v)$ is not differentiable in one direction. 
If $E_0$ were differentiable, for instance if it were smeared out a tiny bit by
convolution with a narrow Gaussian, it would display a usual saddle point very close
to $(v_0/4,v_0)$. We interprete the occurence of this special point on 
the line $(v/4,v)$ as evidence that the optimum $\mu\k$ should not 
display a finite dispersion on the boundary of the magnetic Brillouin zone,
cf.\ Eq.\ \eqref{eq:boundary_dispersion}.

\begin{figure}[htb]
\centering
\includegraphics[width=0.98\columnwidth,clip]{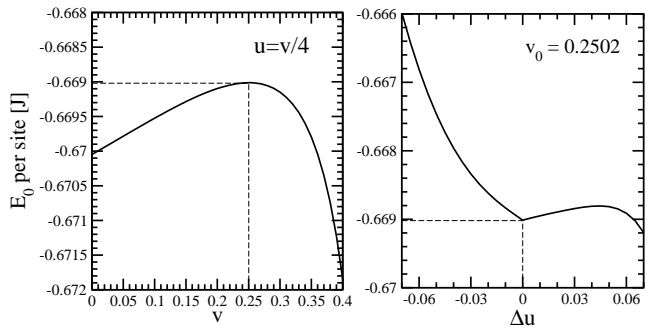}
\caption{\label{fig:energy_cuts} 
Cuts of $E_0(u,v)/(2N)$ along special lines. Left panel: along the lines
of cusps $u=v/4$. Clearly a local maximum appears at $v_0=0.2502$.
Right panel: This cut follows the line through the point $(u_0=v_0/4,v_0)$,
but perpendicular to $u=v/4$. It is given by $(u_0+\Delta u,v_0-\Delta u/4)$.
The cusp at $\Delta u=0$ is obvious.}
\end{figure}

For the precise determination of $v_0$, calculations are done for various
system sizes with linear extensions $L=24, 32, 36$. The extrapolation of the
position of the local maximum yields $v_0=0.2502(1)$ and the energy value at
this position is found to be $E_0/(2N)=-0.66902(2)J$.
These values should be compared to the quantum Monte Carlo result \cite{sandv97b}
$E_0/(2N)=-0.669437(5)J$  and to the second order result
of a plain $1/S$ expansion \cite{hamer92} which reads $E_0/(2N)=-0.66999J$.
(Note that this number is referred to as ``third order'' in Ref.\ 
\onlinecite{hamer92} because the authors include the classical energy in their
power counting.) If we take the Monte Carlo expansion as reliable reference
the variation of second order perturbation theory
could reduce the deviation from $0.0006J$ to $-0.0004J$
which is a reduction by about 25\%.
To judge the improvement we point out that
passing from rather simple first order perturbative spin wave
theory  $E_0/(2N)=-0.67042J$ to second order $E_0/(2N)=-0.66999J$ 
improved the ground state energy by 44\%.

We also stress that the improved result $E_0/(2N)=-0.66902(2)J$ 
is obtained by  using equations of the same complexity as the second order equations.
The add-on is just the variation of the unperturbed starting point.
In higher orders, this variation becomes an even more efficient tool,
see Ref.\ \onlinecite{dusue04a} for the discussion of the example of 
the quartic oscillator.

In the end, however, our goal is not to obtain estimates for the ground
state energy in the first place. In the varied perturbation theory,
the analysis of the ground state energy primarily serves the purpose 
to fix the unperturbed starting point $H_D$.

\subsection{\label{sec:dispersion}Magnon Dispersion}

Above, we have determined the optimum starting point $H_D(u,v)$
by analysing the dependence of the approximate ground state energy.
We identified the optimum starting point to be given by
$(v_0/4,v_0)$ with $v_0=0.2502(1)$ where a saddle point
appeared. Next, we use this starting point to analyze the
magnon dispersion in general and the dip between the values
at $(\pi,0)$ and at $(\pi/2,\pi/2)$ in particular.

\begin{figure}[htb]
\centering
\includegraphics[width=0.96\columnwidth,clip]{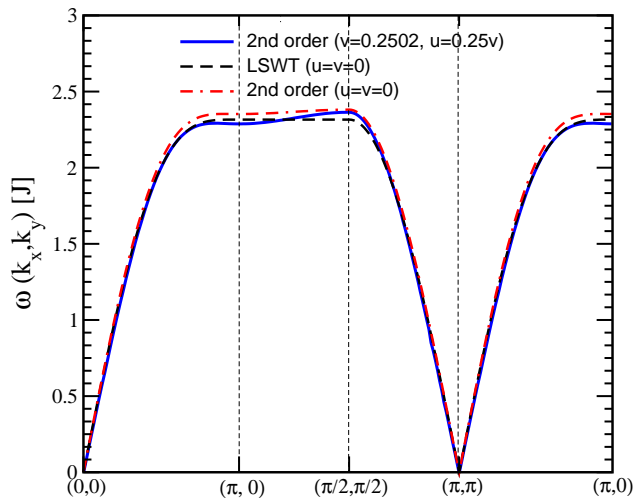}
\caption{\label{fig:dispersion} 
(Color online) Dispersion of the Heisenberg lattice model in (i) linear spin
wave theory (LSWT) including first order corrections,
(ii) second order perturbation theory around the LSWT solution, i.e., 
for $(u,v)=(0,0)$, (iii) in varied second order perturbation theory
around $(v_0/4,v_0)$ with $v_0=0.2502$ which corresponds to the saddle point
of the ground state energy.}
\end{figure}

Fig.\ \ref{fig:dispersion} depicts the corresponding result compared with
results from first and second order perturbative spin wave theory.
First, we find that the overall shape of all three curves is very similar.
This was expected from the comparison of various perturbative results, high
order series expansion and quantum Monte Carlo data, see Fig.\ 1 in Ref.\
\onlinecite{syrom10}.

Second, the dip at $(\pi,0)$ relative to the dispersion at $(\pi/2,\pi/2)$
is more pronounced in the varied perturbation theory. We find that the
dip takes the relative value $3.3(1)\%$ which is rather precisely the value
which Syromyatnikov found in the much more complex third order perturbation
calculation \cite{syrom10}. It improves the second order result of $1.4\%$
by more than a factor of 2 while it is still away from the about $9\%$ of dip
obtained by series expansion \cite{zheng05} or quantum Monte Carlo \cite{sandv01}.

In detail, we find $\omega(\pi,0) = 2.2881(1)J$ and $\omega(\pi/2,\pi/2) = 2.3643(2)J$.
The latter value is very close to the series value $\omega^\text{series}(\pi/2,\pi/2) = 2.385(1)J$
and to the quantum Monte Carlo value $\omega^\text{QMC}(\pi/2,\pi/2) = 2.39J$.
The former is still by about 5\% too high compared to $\omega^\text{series}(\pi,0) = 2.18(1)J$
and  $\omega^\text{QMC}(\pi,0) = 2.16J$. So there is still some way to go.

But to judge the numbers obtained by varied perturbation theory 
we also compare to the ordinary second order perturbative numbers
$\omega^\text{2nd}(\pi,0)=2.3586J$ and $\omega^\text{2nd}(\pi/2,\pi/2)=2.3920J$
and to the third order perturbative numbers 
$\omega^\text{3rd}(\pi,0)=2.3241(2)J$ and $\omega^\text{3rd}(\pi/2,\pi/2)=2.4007(2)J$.
Relative to these numbers, the varied perturbative results represent an improvement,
in particular in comparison to the plain second order results.
It must be kept in mind that the varied perturbation theory is based essentially
on the same equations as the plain second order results.
Thus one achieves third order accuracy, see results by Syromyatnikov in Ref.\ \onlinecite{syrom10},
for the effort of the second order calculation.

These findings show that the variation of perturbative calculations 
indeed reduces deviations to the exact results. In this way,  
improved results are accessible without resorting to more complex higher order
calculations.

\section{\label{sec:conclusio}Conclusions}

In summary, we investigated the Heisenberg quantum antiferromagnet in terms
of spin waves (magnons) based on the Dyson-Maleev representation.
The primary goal was to determine the dispersion of the magnons. 
A secondary goal was the determination of the ground state energy.

The approach employed for evaluation is based on standard perturbation theory.
But we do not pursue a plain expansion in $1/S$. Instead, we choose
the unperturbed Hamiltonian $H_D$, which serves as starting point,
arbitrarily among bilinear bosonic operators. In the present article,
we did not exploit the full freedom of choice of such operators but
investigated a parametrization with two variables $(u,v)$
which remains close to the bilinear
bosonic Hamiltonian of linear spin wave theory. Considering
more variables would have obscured the fundamental principle
of the approach and it would have been rather cumbersome on the technical level.

Following the principle of minimal sensitivity, we search the parameter
space $(u,v)$ for stationary points, i.e., local extrema or saddle points,
 of $E_0(u,v)$. Thus the approach is called varied perturbation theory.
 Such a point is indeed found at $(u_0=v_0/4,v_0)$ with
 $v_0=0.2502(1)$. Due to non-differentiability, it is not an ordinary 
 saddle point, but a point with a cusp-like minimum in one direction and
 a differentiable maximum in the perpendicular direction.
   
At this saddle point, the value of the ground state energy is closer to the
exact value than the plain second order perturbation. 
Furthermore, the magnon dispersion obtained at this saddle point
displays a more significant dip of $3.3\%$ which is again more than
twice as large as the plain second order calculation provides.
The plain third order calculation yields a comparable dip
of $3.2\%$ which means that the variation of the starting point
allowed us to obtain third order accuracy with the effort of
a second order calculation. This represents the methodological
achievement.
The agreement with high order series results and quantum Monte Carlo
data is still unsatisfactory because these approaches find 
a dip of about $9\%$. 

We attribute the remaining discrepancy 
to the low order (here: second order) perturbative approach
which we employed to calculate $E_0(u,v)$. We expect that 
a varied third order approach enhances the dip to about  $6 - 7\%$ 
percent, combining the factors of 2 from the variation (this article) and from
passing from second to third order  (Ref.\ \onlinecite{syrom10}).

Previous investigations of the simple model of a quartic oscillator
have shown that the variation of the starting point combined with improved
evaluation schemes such as higher order perturbation theory \cite{steve81}
or continuous unitary transformation \cite{dusue04a} is capable to
provide very good quantitative results.

The progress achieved in this article is two-fold:
On the methodological side, we introduced the principle
of minimal sensitivity in the calculation for an 
examplary extended solid state system displaying
important correlations.

On the physical side, we provided evidence that the
dip in the dispersion of the square lattice
Heisenberg antiferromagnet with $S=1/2$ can be
explained in terms of magnons if more advanced
approaches are used. In our opinion, one does not
have to resort to fractionalization into spinons
as sometimes discussed \cite{chris07} in order to 
understand the dip.

But we admit that a quantitative
reproduction of the dip has not yet been achieved so that
further work is called for. Promising improved approaches
to reach this goal
comprise third order perturbative approaches and
continuous unitary transformations \cite{krull12}.

\begin{acknowledgments}
We acknowledge the Texas Advanced Computing Center (TACC) at 
The University of Texas at Austin for providing HPC resources 
that have contributed to part of the research results reported within this paper. 
Part of this work was also done at the HPC cluster at GVSU which is supported by 
the National Science Foundation  Grant No. CNS-1228291.
We acknowledge financial support by the 
NRW Forschungsschule ``Forschung mit Synchrotronstrahlung 
in den Nano- und Biowissenschaften'' and by the
the Helmholtz Virtual Institute 
``New states of matter and their excitations''.
\end{acknowledgments}

\appendix

\section{Vertex Functions}
\label{sec:vertex}

Below, we use $x_i$ for $x_{{\bf k}_i}$ and $\gamma(i)$ for 
$\gamma({\bf k}_i)$, $\gamma(i-j)$ for $\gamma({\bf k}_i- {\bf k}_j)$,
and so on. The vertex functions are given by
\begin{widetext}
\begin{subequations}
\bea
V^{(1)}_{1234} &=& x_1 x_3\gamma(1-3)+x_1x_4\gamma(1-4)+x_2x_3\gamma(2-3)
+x_2x_4 \gamma(2-4) \non \\
&&- x_1\gamma(1)-x_2\gamma(2)-x_1x_3x_4\gamma(1-3-4)
-x_2x_3x_4\gamma(2-3-4),
\\
V^{(2)}_{1234} &=& -x_3\gamma(2-3)-x_4\gamma(2-4)-x_1x_2x_3\gamma(1-3)
-x_1x_2x_4 \gamma(1-4) \non \\
&&+ x_1x_2\gamma(1)+\gamma(2)+x_1x_2x_3x_4\gamma(1-3-4)
+x_3x_4\gamma(2-3-4),
\\
V^{(3)}_{1234} &=& -x_1\gamma(1-3)-x_2\gamma(2-3)-x_1x_3x_4\gamma(1-4)
-x_2x_3x_4 \gamma(2-4) \non \\
&&+ x_1x_3\gamma(1)+x_2x_3\gamma(2)+x_1x_4\gamma(1-3-4)
+x_2x_4\gamma(2-3-4), 
\\
V^{(4)}_{1234} &=& x_1x_2x_3x_4\gamma(1-3)+x_1x_2\gamma(1-4)+x_3x_4\gamma(2-3)
+\gamma(2-4) \non \\
&&- x_4\gamma(2)-x_1x_2x_4\gamma(1)-x_3\gamma(2-3-4)
-x_1x_2x_3\gamma(1-3-4),
\\
V^{(5)}_{1234} &=& -x_2x_3x_4\gamma(1-3)-x_1x_3x_4\gamma(2-3)-x_1\gamma(2-4)
-x_2\gamma(1-4) \non \\
&&+ x_1x_4\gamma(2)+x_2x_4\gamma(1)+x_1x_3\gamma(2-3-4)
+x_2x_3\gamma(1-3-4),
\\
V^{(6)}_{1234} &=& -x_{4}\gamma(1-3)-x_3\gamma(1-4)-x_1x_2x_3\gamma(2-4)
-x_1x_2x_4\gamma(2-3) \non \\
&&+\gamma(1-3-4)+x_1x_2\gamma(2-3-4)+x_3x_4\gamma(1)
+x_1x_2x_3x_4\gamma(2),
\\
V^{(7)}_{1234} &=& x_1x_4\gamma(1-3)+x_1x_3\gamma(1-4)+x_2x_3\gamma(2-4)
+x_2x_4\gamma(2-3) \non \\
&&- x_1x_3x_4\gamma(1)-x_2x_3x_4\gamma(2)-x_1\gamma(1-3-4)
-x_2\gamma(2-3-4),
\\
V^{(8)}_{1234} &=& x_1x_4\gamma(2-4)+x_2x_4\gamma(1-4)+x_1x_3\gamma(2-3)
+x_2x_3\gamma(1-3) \non \\
&&- x_1\gamma(2)-x_2\gamma(1)-x_1x_3x_4\gamma(2-3-4)
-x_2x_3x_4\gamma(1-3-4),
\\
V^{(9)}_{1234} &=& x_1x_3\gamma(2-4)+x_2x_3\gamma(1-4)+x_1x_4\gamma(2-3)
+x_2x_4\gamma(1-3) \non \\
&&- x_1\gamma(2-3-4)-x_2\gamma(1-3-4)-x_1x_3x_4\gamma(2)
-x_2x_3x_4\gamma(1).
\eea
\end{subequations}
\end{widetext}


\begin{thebibliography}{26}%
\makeatletter
\providecommand \@ifxundefined [1]{%
 \@ifx{#1\undefined}
}%
\providecommand \@ifnum [1]{%
 \ifnum #1\expandafter \@firstoftwo
 \else \expandafter \@secondoftwo
 \fi
}%
\providecommand \@ifx [1]{%
 \ifx #1\expandafter \@firstoftwo
 \else \expandafter \@secondoftwo
 \fi
}%
\providecommand \natexlab [1]{#1}%
\providecommand \enquote  [1]{``#1''}%
\providecommand \bibnamefont  [1]{#1}%
\providecommand \bibfnamefont [1]{#1}%
\providecommand \citenamefont [1]{#1}%
\providecommand \href@noop [0]{\@secondoftwo}%
\providecommand \href [0]{\begingroup \@sanitize@url \@href}%
\providecommand \@href[1]{\@@startlink{#1}\@@href}%
\providecommand \@@href[1]{\endgroup#1\@@endlink}%
\providecommand \@sanitize@url [0]{\catcode `\\12\catcode `\$12\catcode
  `\&12\catcode `\#12\catcode `\^12\catcode `\_12\catcode `\%12\relax}%
\providecommand \@@startlink[1]{}%
\providecommand \@@endlink[0]{}%
\providecommand \url  [0]{\begingroup\@sanitize@url \@url }%
\providecommand \@url [1]{\endgroup\@href {#1}{\urlprefix }}%
\providecommand \urlprefix  [0]{URL }%
\providecommand \Eprint [0]{\href }%
\providecommand \doibase [0]{http://dx.doi.org/}%
\providecommand \selectlanguage [0]{\@gobble}%
\providecommand \bibinfo  [0]{\@secondoftwo}%
\providecommand \bibfield  [0]{\@secondoftwo}%
\providecommand \translation [1]{[#1]}%
\providecommand \BibitemOpen [0]{}%
\providecommand \bibitemStop [0]{}%
\providecommand \bibitemNoStop [0]{.\EOS\space}%
\providecommand \EOS [0]{\spacefactor3000\relax}%
\providecommand \BibitemShut  [1]{\csname bibitem#1\endcsname}%
\let\auto@bib@innerbib\@empty
\bibitem [{\citenamefont {Manousakis}(1991)}]{manou91}%
  \BibitemOpen
  \bibfield  {author} {\bibinfo {author} {\bibfnamefont {E.}~\bibnamefont
  {Manousakis}},\ }\href@noop {} {\bibfield  {journal} {\bibinfo  {journal}
  {Rev. Mod. Phys.}\ }\textbf {\bibinfo {volume} {63}},\ \bibinfo {pages} {1}
  (\bibinfo {year} {1991})}\BibitemShut {NoStop}%
\bibitem [{\citenamefont {Majumdar}\ \emph {et~al.}(2012)\citenamefont
  {Majumdar}, \citenamefont {Furton},\ and\ \citenamefont {Uhrig}}]{majum12a}%
  \BibitemOpen
  \bibfield  {author} {\bibinfo {author} {\bibfnamefont {K.}~\bibnamefont
  {Majumdar}}, \bibinfo {author} {\bibfnamefont {D.}~\bibnamefont {Furton}}, \
  and\ \bibinfo {author} {\bibfnamefont {G.~S.}\ \bibnamefont {Uhrig}},\
  }\href@noop {} {\bibfield  {journal} {\bibinfo  {journal} {Phys. Rev. B}\
  }\textbf {\bibinfo {volume} {85}},\ \bibinfo {pages} {144420} (\bibinfo
  {year} {2012})}\BibitemShut {NoStop}%
\bibitem [{\citenamefont {\protect{Le Tacon}}\ \emph
  {et~al.}(2011)\citenamefont {\protect{Le Tacon}}, \citenamefont
  {Ghiringhelli}, \citenamefont {Chaloupka}, \citenamefont {Sala},
  \citenamefont {Hinkov}, \citenamefont {Haverkort}, \citenamefont {Minola},
  \citenamefont {Bakr}, \citenamefont {Zhou}, \citenamefont {Blanco-Canosa},
  \citenamefont {Monney}, \citenamefont {Song}, \citenamefont {Sun},
  \citenamefont {Lin}, \citenamefont {Luca}, \citenamefont {Salluzzo},
  \citenamefont {Khaliullin}, \citenamefont {Schmitt}, \citenamefont
  {Braicovic},\ and\ \citenamefont {Keimer}}]{letac11}%
  \BibitemOpen
  \bibfield  {author} {\bibinfo {author} {\bibfnamefont {M.}~\bibnamefont
  {\protect{Le Tacon}}}, \bibinfo {author} {\bibfnamefont {G.}~\bibnamefont
  {Ghiringhelli}}, \bibinfo {author} {\bibfnamefont {J.}~\bibnamefont
  {Chaloupka}}, \bibinfo {author} {\bibfnamefont {M.~M.}\ \bibnamefont {Sala}},
  \bibinfo {author} {\bibfnamefont {V.}~\bibnamefont {Hinkov}}, \bibinfo
  {author} {\bibfnamefont {M.}~\bibnamefont {Haverkort}}, \bibinfo {author}
  {\bibfnamefont {M.}~\bibnamefont {Minola}}, \bibinfo {author} {\bibfnamefont
  {M.}~\bibnamefont {Bakr}}, \bibinfo {author} {\bibfnamefont {K.~J.}\
  \bibnamefont {Zhou}}, \bibinfo {author} {\bibfnamefont {S.}~\bibnamefont
  {Blanco-Canosa}}, \bibinfo {author} {\bibfnamefont {C.}~\bibnamefont
  {Monney}}, \bibinfo {author} {\bibfnamefont {Y.~T.}\ \bibnamefont {Song}},
  \bibinfo {author} {\bibfnamefont {G.~L.}\ \bibnamefont {Sun}}, \bibinfo
  {author} {\bibfnamefont {C.~T.}\ \bibnamefont {Lin}}, \bibinfo {author}
  {\bibfnamefont {G.~M.~D.}\ \bibnamefont {Luca}}, \bibinfo {author}
  {\bibfnamefont {M.}~\bibnamefont {Salluzzo}}, \bibinfo {author}
  {\bibfnamefont {G.}~\bibnamefont {Khaliullin}}, \bibinfo {author}
  {\bibfnamefont {T.}~\bibnamefont {Schmitt}}, \bibinfo {author} {\bibfnamefont
  {L.}~\bibnamefont {Braicovic}}, \ and\ \bibinfo {author} {\bibfnamefont
  {B.}~\bibnamefont {Keimer}},\ }\href@noop {} {\bibfield  {journal} {\bibinfo
  {journal} {Nature Phys.}\ }\textbf {\bibinfo {volume} {7}},\ \bibinfo {pages}
  {725} (\bibinfo {year} {2011})}\BibitemShut {NoStop}%
\bibitem [{\citenamefont {Singh}\ and\ \citenamefont
  {Gelfand}(1995)}]{singh95}%
  \BibitemOpen
  \bibfield  {author} {\bibinfo {author} {\bibfnamefont {R.~R.~P.}\
  \bibnamefont {Singh}}\ and\ \bibinfo {author} {\bibfnamefont {M.~P.}\
  \bibnamefont {Gelfand}},\ }\href@noop {} {\bibfield  {journal} {\bibinfo
  {journal} {Phys. Rev. B}\ }\textbf {\bibinfo {volume} {52}},\ \bibinfo
  {pages} {15695} (\bibinfo {year} {1995})}\BibitemShut {NoStop}%
\bibitem [{\citenamefont {Zheng}\ \emph {et~al.}(2005)\citenamefont {Zheng},
  \citenamefont {Oitmaa},\ and\ \citenamefont {Hamer}}]{zheng05}%
  \BibitemOpen
  \bibfield  {author} {\bibinfo {author} {\bibfnamefont {W.}~\bibnamefont
  {Zheng}}, \bibinfo {author} {\bibfnamefont {J.}~\bibnamefont {Oitmaa}}, \
  and\ \bibinfo {author} {\bibfnamefont {C.~J.}\ \bibnamefont {Hamer}},\
  }\href@noop {} {\bibfield  {journal} {\bibinfo  {journal} {Phys. Rev. B}\
  }\textbf {\bibinfo {volume} {71}},\ \bibinfo {pages} {184440} (\bibinfo
  {year} {2005})}\BibitemShut {NoStop}%
\bibitem [{\citenamefont {Sandvik}\ and\ \citenamefont
  {Singh}(2001)}]{sandv01}%
  \BibitemOpen
  \bibfield  {author} {\bibinfo {author} {\bibfnamefont {A.~W.}\ \bibnamefont
  {Sandvik}}\ and\ \bibinfo {author} {\bibfnamefont {R.~R.~P.}\ \bibnamefont
  {Singh}},\ }\href@noop {} {\bibfield  {journal} {\bibinfo  {journal} {Phys.
  Rev. Lett.}\ }\textbf {\bibinfo {volume} {86}},\ \bibinfo {pages} {528}
  (\bibinfo {year} {2001})}\BibitemShut {NoStop}%
\bibitem [{\citenamefont {Igarashi}(1992)}]{igara92}%
  \BibitemOpen
  \bibfield  {author} {\bibinfo {author} {\bibfnamefont {J.}~\bibnamefont
  {Igarashi}},\ }\href@noop {} {\bibfield  {journal} {\bibinfo  {journal}
  {Phys. Rev. Lett.}\ }\textbf {\bibinfo {volume} {46}},\ \bibinfo {pages}
  {10763} (\bibinfo {year} {1992})}\BibitemShut {NoStop}%
\bibitem [{\citenamefont {Hamer}\ \emph {et~al.}(1992)\citenamefont {Hamer},
  \citenamefont {Zheng},\ and\ \citenamefont {Arndt}}]{hamer92}%
  \BibitemOpen
  \bibfield  {author} {\bibinfo {author} {\bibfnamefont {C.~J.}\ \bibnamefont
  {Hamer}}, \bibinfo {author} {\bibfnamefont {W.}~\bibnamefont {Zheng}}, \ and\
  \bibinfo {author} {\bibfnamefont {P.}~\bibnamefont {Arndt}},\ }\href@noop {}
  {\bibfield  {journal} {\bibinfo  {journal} {Phys. Rev. B}\ }\textbf {\bibinfo
  {volume} {46}},\ \bibinfo {pages} {6276} (\bibinfo {year}
  {1992})}\BibitemShut {NoStop}%
\bibitem [{\citenamefont {Igarashi}\ and\ \citenamefont
  {Nagao}(2005)}]{igara05}%
  \BibitemOpen
  \bibfield  {author} {\bibinfo {author} {\bibfnamefont {J.}~\bibnamefont
  {Igarashi}}\ and\ \bibinfo {author} {\bibfnamefont {T.}~\bibnamefont
  {Nagao}},\ }\href@noop {} {\bibfield  {journal} {\bibinfo  {journal} {Phys.
  Rev. B}\ }\textbf {\bibinfo {volume} {72}},\ \bibinfo {pages} {014403}
  (\bibinfo {year} {2005})}\BibitemShut {NoStop}%
\bibitem [{\citenamefont {Syromyatnikov}(2010)}]{syrom10}%
  \BibitemOpen
  \bibfield  {author} {\bibinfo {author} {\bibfnamefont {A.~V.}\ \bibnamefont
  {Syromyatnikov}},\ }\href@noop {} {\bibfield  {journal} {\bibinfo  {journal}
  {J. Phys. C}\ }\textbf {\bibinfo {volume} {22}},\ \bibinfo {pages} {216003}
  (\bibinfo {year} {2010})}\BibitemShut {NoStop}%
\bibitem [{\citenamefont {R\protect{\o}nnow}\ \emph {et~al.}(2001)\citenamefont
  {R\protect{\o}nnow}, \citenamefont {McMorrow}, \citenamefont {Coldea},
  \citenamefont {Harrison}, \citenamefont {Youngson}, \citenamefont {Perring},
  \citenamefont {Aeppli}, \citenamefont {Sylju\protect{\aa}sen}, \citenamefont
  {Lefmann},\ and\ \citenamefont {Rischel}}]{ronno01}%
  \BibitemOpen
  \bibfield  {author} {\bibinfo {author} {\bibfnamefont {H.~M.}\ \bibnamefont
  {R\protect{\o}nnow}}, \bibinfo {author} {\bibfnamefont {D.~F.}\ \bibnamefont
  {McMorrow}}, \bibinfo {author} {\bibfnamefont {R.}~\bibnamefont {Coldea}},
  \bibinfo {author} {\bibfnamefont {A.}~\bibnamefont {Harrison}}, \bibinfo
  {author} {\bibfnamefont {I.~D.}\ \bibnamefont {Youngson}}, \bibinfo {author}
  {\bibfnamefont {T.~G.}\ \bibnamefont {Perring}}, \bibinfo {author}
  {\bibfnamefont {G.}~\bibnamefont {Aeppli}}, \bibinfo {author} {\bibfnamefont
  {O.}~\bibnamefont {Sylju\protect{\aa}sen}}, \bibinfo {author} {\bibfnamefont
  {K.}~\bibnamefont {Lefmann}}, \ and\ \bibinfo {author} {\bibfnamefont
  {C.}~\bibnamefont {Rischel}},\ }\href@noop {} {\bibfield  {journal} {\bibinfo
   {journal} {Phys. Rev. Lett.}\ }\textbf {\bibinfo {volume} {87}},\ \bibinfo
  {pages} {037202} (\bibinfo {year} {2001})}\BibitemShut {NoStop}%
\bibitem [{\citenamefont {Christensen}\ \emph {et~al.}(2004)\citenamefont
  {Christensen}, \citenamefont {McMorrow}, \citenamefont {R\o{}nnow},
  \citenamefont {Harrison}, \citenamefont {Perring},\ and\ \citenamefont
  {Coldea}}]{chris04b}%
  \BibitemOpen
  \bibfield  {author} {\bibinfo {author} {\bibfnamefont {N.~B.}\ \bibnamefont
  {Christensen}}, \bibinfo {author} {\bibfnamefont {D.~F.}\ \bibnamefont
  {McMorrow}}, \bibinfo {author} {\bibfnamefont {H.~M.}\ \bibnamefont
  {R\o{}nnow}}, \bibinfo {author} {\bibfnamefont {A.}~\bibnamefont {Harrison}},
  \bibinfo {author} {\bibfnamefont {T.~G.}\ \bibnamefont {Perring}}, \ and\
  \bibinfo {author} {\bibfnamefont {R.}~\bibnamefont {Coldea}},\ }\href@noop {}
  {\bibfield  {journal} {\bibinfo  {journal} {J. Mag. Mag. Mat.}\ }\textbf
  {\bibinfo {volume} {272-276}},\ \bibinfo {pages} {896} (\bibinfo {year}
  {2004})}\BibitemShut {NoStop}%
\bibitem [{\citenamefont {Christensen}\ \emph {et~al.}(2007)\citenamefont
  {Christensen}, \citenamefont {R\o{}nnow}, \citenamefont {McMorrow},
  \citenamefont {Harrison}, \citenamefont {Perring}, \citenamefont {Enderle},
  \citenamefont {Coldea}, \citenamefont {Regnault},\ and\ \citenamefont
  {Aeppli}}]{chris07}%
  \BibitemOpen
  \bibfield  {author} {\bibinfo {author} {\bibfnamefont {N.~B.}\ \bibnamefont
  {Christensen}}, \bibinfo {author} {\bibfnamefont {H.~M.}\ \bibnamefont
  {R\o{}nnow}}, \bibinfo {author} {\bibfnamefont {D.~F.}\ \bibnamefont
  {McMorrow}}, \bibinfo {author} {\bibfnamefont {A.}~\bibnamefont {Harrison}},
  \bibinfo {author} {\bibfnamefont {T.~G.}\ \bibnamefont {Perring}}, \bibinfo
  {author} {\bibfnamefont {M.}~\bibnamefont {Enderle}}, \bibinfo {author}
  {\bibfnamefont {R.}~\bibnamefont {Coldea}}, \bibinfo {author} {\bibfnamefont
  {L.~P.}\ \bibnamefont {Regnault}}, \ and\ \bibinfo {author} {\bibfnamefont
  {G.}~\bibnamefont {Aeppli}},\ }\href@noop {} {\bibfield  {journal} {\bibinfo
  {journal} {Proc. Nat. Acad. Sciences}\ }\textbf {\bibinfo {volume} {104}},\
  \bibinfo {pages} {15264} (\bibinfo {year} {2007})}\BibitemShut {NoStop}%
\bibitem [{\citenamefont {Headings}\ \emph {et~al.}(2010)\citenamefont
  {Headings}, \citenamefont {Hayden}, \citenamefont {Coldea},\ and\
  \citenamefont {Perring}}]{headi10}%
  \BibitemOpen
  \bibfield  {author} {\bibinfo {author} {\bibfnamefont {N.~S.}\ \bibnamefont
  {Headings}}, \bibinfo {author} {\bibfnamefont {S.~M.}\ \bibnamefont
  {Hayden}}, \bibinfo {author} {\bibfnamefont {R.}~\bibnamefont {Coldea}}, \
  and\ \bibinfo {author} {\bibfnamefont {T.~G.}\ \bibnamefont {Perring}},\
  }\href@noop {} {\bibfield  {journal} {\bibinfo  {journal} {Phys. Rev. Lett.}\
  }\textbf {\bibinfo {volume} {105}},\ \bibinfo {pages} {247001} (\bibinfo
  {year} {2010})}\BibitemShut {NoStop}%
\bibitem [{\citenamefont {Knetter}\ \emph {et~al.}(2001)\citenamefont
  {Knetter}, \citenamefont {Schmidt}, \citenamefont {Gr\"uninger},\ and\
  \citenamefont {Uhrig}}]{knett01b}%
  \BibitemOpen
  \bibfield  {author} {\bibinfo {author} {\bibfnamefont {C.}~\bibnamefont
  {Knetter}}, \bibinfo {author} {\bibfnamefont {K.~P.}\ \bibnamefont
  {Schmidt}}, \bibinfo {author} {\bibfnamefont {M.}~\bibnamefont
  {Gr\"uninger}}, \ and\ \bibinfo {author} {\bibfnamefont {G.~S.}\ \bibnamefont
  {Uhrig}},\ }\href@noop {} {\bibfield  {journal} {\bibinfo  {journal} {Phys.
  Rev. Lett.}\ }\textbf {\bibinfo {volume} {87}},\ \bibinfo {pages} {167204}
  (\bibinfo {year} {2001})}\BibitemShut {NoStop}%
\bibitem [{\citenamefont {Schmidt}\ and\ \citenamefont
  {Uhrig}(2005)}]{schmi05b}%
  \BibitemOpen
  \bibfield  {author} {\bibinfo {author} {\bibfnamefont {K.~P.}\ \bibnamefont
  {Schmidt}}\ and\ \bibinfo {author} {\bibfnamefont {G.~S.}\ \bibnamefont
  {Uhrig}},\ }\href@noop {} {\bibfield  {journal} {\bibinfo  {journal} {Mod.
  Phys. Lett. B}\ }\textbf {\bibinfo {volume} {19}},\ \bibinfo {pages} {1179}
  (\bibinfo {year} {2005})}\BibitemShut {NoStop}%
\bibitem [{\citenamefont {Knetter}\ and\ \citenamefont
  {Uhrig}(2000)}]{knett00a}%
  \BibitemOpen
  \bibfield  {author} {\bibinfo {author} {\bibfnamefont {C.}~\bibnamefont
  {Knetter}}\ and\ \bibinfo {author} {\bibfnamefont {G.~S.}\ \bibnamefont
  {Uhrig}},\ }\href@noop {} {\bibfield  {journal} {\bibinfo  {journal} {Eur.
  Phys. J. B}\ }\textbf {\bibinfo {volume} {13}},\ \bibinfo {pages} {209}
  (\bibinfo {year} {2000})}\BibitemShut {NoStop}%
\bibitem [{\citenamefont {Trebst}\ \emph {et~al.}(2000)\citenamefont {Trebst},
  \citenamefont {Monien}, \citenamefont {Hamer}, \citenamefont {Weihong},\ and\
  \citenamefont {Singh}}]{trebs00}%
  \BibitemOpen
  \bibfield  {author} {\bibinfo {author} {\bibfnamefont {S.}~\bibnamefont
  {Trebst}}, \bibinfo {author} {\bibfnamefont {H.}~\bibnamefont {Monien}},
  \bibinfo {author} {\bibfnamefont {C.~J.}\ \bibnamefont {Hamer}}, \bibinfo
  {author} {\bibfnamefont {Z.}~\bibnamefont {Weihong}}, \ and\ \bibinfo
  {author} {\bibfnamefont {R.~R.~P.}\ \bibnamefont {Singh}},\ }\href@noop {}
  {\bibfield  {journal} {\bibinfo  {journal} {Phys. Rev. Lett.}\ }\textbf
  {\bibinfo {volume} {85}},\ \bibinfo {pages} {4373} (\bibinfo {year}
  {2000})}\BibitemShut {NoStop}%
\bibitem [{\citenamefont {Stevenson}(1981)}]{steve81}%
  \BibitemOpen
  \bibfield  {author} {\bibinfo {author} {\bibfnamefont {P.~M.}\ \bibnamefont
  {Stevenson}},\ }\href@noop {} {\bibfield  {journal} {\bibinfo  {journal}
  {Phys. Rev. D}\ }\textbf {\bibinfo {volume} {23}},\ \bibinfo {pages} {2916}
  (\bibinfo {year} {1981})}\BibitemShut {NoStop}%
\bibitem [{\citenamefont {Dusuel}\ and\ \citenamefont
  {Uhrig}(2004)}]{dusue04a}%
  \BibitemOpen
  \bibfield  {author} {\bibinfo {author} {\bibfnamefont {S.}~\bibnamefont
  {Dusuel}}\ and\ \bibinfo {author} {\bibfnamefont {G.~S.}\ \bibnamefont
  {Uhrig}},\ }\href@noop {} {\bibfield  {journal} {\bibinfo  {journal} {J.
  Phys. A: Math. Gen.}\ }\textbf {\bibinfo {volume} {37}},\ \bibinfo {pages}
  {9275} (\bibinfo {year} {2004})}\BibitemShut {NoStop}%
\bibitem [{\citenamefont {Dyson}(1956{\natexlab{a}})}]{dyson56a}%
  \BibitemOpen
  \bibfield  {author} {\bibinfo {author} {\bibfnamefont {F.~J.}\ \bibnamefont
  {Dyson}},\ }\href@noop {} {\bibfield  {journal} {\bibinfo  {journal} {Phys.
  Rev.}\ }\textbf {\bibinfo {volume} {102}},\ \bibinfo {pages} {1217} (\bibinfo
  {year} {1956}{\natexlab{a}})}\BibitemShut {NoStop}%
\bibitem [{\citenamefont {Dyson}(1956{\natexlab{b}})}]{dyson56b}%
  \BibitemOpen
  \bibfield  {author} {\bibinfo {author} {\bibfnamefont {F.~J.}\ \bibnamefont
  {Dyson}},\ }\href@noop {} {\bibfield  {journal} {\bibinfo  {journal} {Phys.
  Rev.}\ }\textbf {\bibinfo {volume} {102}},\ \bibinfo {pages} {1230} (\bibinfo
  {year} {1956}{\natexlab{b}})}\BibitemShut {NoStop}%
\bibitem [{\citenamefont {Maleev}(1957)}]{malee57}%
  \BibitemOpen
  \bibfield  {author} {\bibinfo {author} {\bibfnamefont {S.~V.}\ \bibnamefont
  {Maleev}},\ }\href@noop {} {\bibfield  {journal} {\bibinfo  {journal} {Zh.
  Eksp. Teor. Fiz.}\ }\textbf {\bibinfo {volume} {33}},\ \bibinfo {pages}
  {1010} (\bibinfo {year} {1957})}\BibitemShut {NoStop}%
\bibitem [{\citenamefont {Maleev}(1958)}]{malee58}%
  \BibitemOpen
  \bibfield  {author} {\bibinfo {author} {\bibfnamefont {S.~V.}\ \bibnamefont
  {Maleev}},\ }\href@noop {} {\bibfield  {journal} {\bibinfo  {journal} {Sov.
  Phys. JETP}\ }\textbf {\bibinfo {volume} {6}},\ \bibinfo {pages} {776}
  (\bibinfo {year} {1958})}\BibitemShut {NoStop}%
\bibitem [{\citenamefont {Sandvik}(1997)}]{sandv97b}%
  \BibitemOpen
  \bibfield  {author} {\bibinfo {author} {\bibfnamefont {A.~W.}\ \bibnamefont
  {Sandvik}},\ }\href@noop {} {\bibfield  {journal} {\bibinfo  {journal} {Phys.
  Rev. B}\ }\textbf {\bibinfo {volume} {56}},\ \bibinfo {pages} {11678}
  (\bibinfo {year} {1997})}\BibitemShut {NoStop}%
\bibitem [{\citenamefont {Krull}\ \emph {et~al.}(2012)\citenamefont {Krull},
  \citenamefont {Drescher},\ and\ \citenamefont {Uhrig}}]{krull12}%
  \BibitemOpen
  \bibfield  {author} {\bibinfo {author} {\bibfnamefont {H.}~\bibnamefont
  {Krull}}, \bibinfo {author} {\bibfnamefont {N.~A.}\ \bibnamefont {Drescher}},
  \ and\ \bibinfo {author} {\bibfnamefont {G.~S.}\ \bibnamefont {Uhrig}},\
  }\href@noop {} {\bibfield  {journal} {\bibinfo  {journal} {Phys. Rev. B}\
  }\textbf {\bibinfo {volume} {86}},\ \bibinfo {pages} {125113} (\bibinfo
  {year} {2012})}\BibitemShut {NoStop}%
\end{thebibliography}

%

\end{document}